\documentclass[11pt,a4paper]{article}
\pdfoutput=1
\usepackage{jheppub}
\usepackage{gensymb}
\usepackage{placeins}
\usepackage{subfigure}
\usepackage{amssymb,amsmath}
\usepackage{graphicx}
\usepackage{color}
\usepackage{cancel}
\usepackage[colorlinks=true
,urlcolor=blue
,citecolor=blue
,linkcolor=blue
,pagecolor=blue
,linktocpage=true
,pdfproducer=medialab
]{hyperref}

 \usepackage[numbers]{natbib}
\usepackage{notoccite}
\makeatletter \renewcommand{\@dotsep}{10000} \makeatother
\def\be{\begin{equation}}
\def\ee{\end{equation}}
\def\bea{\begin{eqnarray}}
\def\eea{\end{eqnarray}}
\def\bi{\begin{itemize}}
\def\ei{\end{itemize}}

\newcommand{\red}[1]{{\color{red} #1}}
\usepackage[nodisplayskipstretch]{setspace}

\newcommand{\mgut}{M_{{\rm GUT}}}
\newcommand{\msusy}{M_{{\rm SUSY}}}
\newcommand{\mLT}{m_{\tilde{L}_{3}}}
\newcommand{\mLFS}{m_{\tilde{L}_{1,2}}}

\DeclareUnicodeCharacter{2212}{-}
\DeclareUnicodeCharacter{202F}{\,}

\begin{document}

\begin{titlepage}
\pagestyle{empty}

\vspace*{0.2in}
\begin{center}
{\Large \bf Third family quasi-Yukawa unification:
Higgsino dark matter, NLSP gluino and all that} \\
\vspace{1cm}
{\bf Qaisar Shafi$^{a,}$\footnote{E-mail: qshafi@udel.edu}, Amit Tiwari$^{a,}$\footnote{E-mail: amitiit@udel.edu} and
Cem Salih $\ddot{\rm U}$n$^{b,c,}$\footnote{E-mail: cemsalihun@uludag.edu.tr}}
\vspace{0.5cm}

{\small \it
$^a$Bartol Research Institute, Department of Physics and Astronomy,
University of Delaware, Newark, DE 19716, USA\\
$^b$Department of Physics, Bursa Uluda\~{g} University, TR16059 Bursa, Turkey \\
$^c$ Departamento de Ciencias Integradas y Centro de Estudios Avanzados en F\'{i}sica Matem\'aticas y Computación, Campus del Carmen, Universidad de Huelva, Huelva 21071, Spain
}

\end{center}

\vspace{0.5cm}
\begin{abstract}

We explore the implications of third family ($t-b-\tau$) quasi-Yukawa unification (QYU) for collider and dark matter (DM) searches within the framework of a supersymmetric $SU(4)_c \times SU(2)_L \times SU(2)_R$ model. The deviation from exact Yukawa unification is quantified through the relation $y_t : y_b : y_\tau = |1+C|:|1-C|:|1+3C|$, with $C$ being a real parameter ($|C| \leq 0.2$). We allow for the breaking of left-right symmetry both by the soft scalar and gaugino mass parameters and obtain a variety of viable solutions that predict the sparticle mass spectrum including LSP DM (whose stability is guaranteed by a $Z_2$ gauge symmetry). We highlight solutions that include an NLSP gluino with mass $\sim$ 1.3-2.5 TeV, which should be accessible at LHC Run 3. There also exist NSLP stop solutions with masses heavier than about 1.8 TeV, which are consistent with the LSP neutralino dark matter relic density through stop-neutralino coannihilation. We identify A-resonance solutions with DM mass $\sim$ 0.8 - 2 TeV, as well as bino-chargino, bino-slepton and bino-stau co-annihilation scenarios. Finally, we also identify Wino-like ($\sim99\%$) and Higgsino-like ($\sim99\%$) solutions whose masses are heavier than about 1.5 TeV and 1 TeV, respectively. These solutions are compatible with the desired dark matter relic density and testable in ongoing and future direct detection experiments.

\end{abstract}
\end{titlepage}


\section{Introduction}
\label{sec:intro}

Low scale supersymmetry (SUSY) remains an attractive extension of the Standard Model (SM) for a number of reasons. Firstly, the gauge hierarchy problem associated with quadratic divergences in the scalar sector of the SM is significantly tamed in the presence of low scale SUSY. Secondly, while the SM quartic Higgs coupling $\lambda$ is essentially a free parameter in the SM, in the simplest supersymmetric extensions such as the Minimal Supersymmetric SM (MSSM), $\lambda$ is related to the gauge couplings of the electroweak sector. This feature allows one to provide an estimate for the upper bound of around 130 GeV or so on the SM Higgs mass in MSSM,  which is in excellent agreement with the experimentally measured value of 125.6 GeV \cite{ATLAS:2012yve,CMS:2013btf}. Thanks to SUSY, the problem of $\lambda$ running to zero and subsequently turning negative at a scale of around $10^{11}$ GeV is also avoided. Finally, in the presence of TeV  scale SUSY, the three SM gauge couplings nicely unify at an energy scale close to $10^{16}$ GeV \cite{Dimopoulos:1981yj,Amaldi:1991cn,Ellis:1990wk,Langacker:1991an}. This last feature provides a strong motivation for considering supersymmetric grand unified theories. Other good reasons include electric charge quantization, unification of quarks and leptons in each family, and prediction of non-zero neutrino masses, which is required by the observed solar, atmospheric and reactor neutrino oscillation experiments (for a recent review and additional references see \cite{Formaggio:2021nfz}). A particularly attractive example of grand unification is provided by SUSY SO(10) which, among other things, also predicts third family ($t-b-\tau$) Yukawa unification (YU) to a good approximation \cite{Ananthanarayan:1991xp,Ananthanarayan:1992cd}. The consequences for collider and dark matter physics that follow from YU have been extensively studied in the literature (see \cite{Gomez:2020gav} for recent discussion and additional references).

Motivated by the ongoing LHC Run 3 at CERN and the large number of dark matter searches underway, we investigate the experimental consequences of third family Quasi YU (QYU) \cite{Dar:2011sj,Shafi:2015lfa,Altin:2017sxx} in the framework of $SU(4)_c \times SU(2)_L \times SU(2)_R$ ($422$ for short), which is a maximal subgroup of $SO(10)$ \cite{Pati:1974yy} and retains some of the key predictions of $SO(10)$ (for a recent discussion of b-tau Yukawa unification in 422 see Ref.~\cite{Ahmed:2022ibc}). We go beyond earlier investigations with the assumption that the soft SUSY breaking scalar and gaugino masses do not respect left-right symmetry, and the soft scalar masses for the first two families are split from the third family which allows us to probe a larger region of the parameter space compared to earlier studies.

We have explored the predictions for sparticle masses including dark matter and NLSP candidates in the framework of a supersymmetric $422$ model which incorporates third family QYU. An unbroken $Z_2$ gauge symmetry contained in the $422$ model acts as matter parity and ensures the presence of a viable neutralino dark matter candidate \cite{Kibble:1982dd}. In addition to Bino-like DM we also identify Wino-like and Higgsino-like dark matter solutions, which yield the desired dark matter relic abundance with masses greater than or of order 1.5 TeV and 1 TeV, respectively. We also identify stop-neutralino coannihilation solutions for masses in the 1.8-2.3 TeV range. The NLSP gluino mass lies in the 1.3-2.5 TeV range, which should be testable at the LHC Run 3. Other solutions include the NLSP stau with mass between 1-2 TeV and A-resonance solutions with the mass $m_A$ varying between 0.5 and 2.5 TeV. In this context, the A-resonance solutions can also be tested through the decay channel A, $H \longrightarrow \tau \tau$, which currently excludes the solutions with $m_A \leq 2$ TeV in the large $\tan\beta$ region. We display several benchmark points that highlight these solutions and also show that the dark matter neutralino may be accessible in the current and near future experiments.

 This paper is organized as follows: In Section \ref{sec:scan}, we briefly describe QYU, scanning procedure, the employed constraints and fundamental parameter space. In section \ref{sec:fundQYU}, we display the plots for the GUT scale mass parameters and the implications for the mass spectra. Section \ref{sec:DM} is devoted to the DM implications including the relic density as well as the spin-independent and spin-dependent scattering cross-sections. In this section, we display five benchmark points and discuss the prospects to test QYU in the ongoing collider and DM experiments.

\section{Quasi-Yukawa Unification, Fundamental Parameters, Scanning Procedure and Experimental Constraints}
\label{sec:scan}

Precise third family Yukawa Unification (YU) is realized in supersymmetric unified theories based on gauge groups such as SO(10) and $422$ \cite{Ananthanarayan:1991xp,Ananthanarayan:1994qt}. Quasi-Yukawa Unification (QYU) is motivated by the desire to incorporate the observed fermion masses and mixings, and a particularly simple yet realistic example of t-b-tau QYU is provided by the relation \cite{Dar:2011sj,Gomez:2002tj}:
\begin{equation}
y_{t}:y_{b}:y_{\tau} = (1+C):(1-C):(1+3C)~,
\label{eq:CQYU}
\end{equation}
where $C$ is taken to be real, but it can be negative or positive. The derivation of this relation in SO(10) and $422$ models can be found in Ref.\cite{Dar:2011sj} and will not be repeated here. Our main goal in this paper is to explore the phenomenological implications of this QYU condition in $422$ models that can be tested in ongoing collider and dark matter experiments. Our investigation has some overlap with earlier work, but an important new ingredient here is the violation of left-right symmetry by the soft SUSY breaking parameters of the scalar and gaugino sectors of the model.

The soft supersymmetry breaking (SSB) terms in the Lagrangian include the mass terms for the scalars and gauginos, as well as trilinear interactions of the supersymmetric particles. However, these SSB terms cannot be arbitrary because of the underlying GUT symmetry. Since we consider a symmetry breaking pattern in which the LR symmetry is also broken, the set of free parameters includes two different mass terms for the left and right handed fields. In addition, we also assume a flavor symmetry at the GUT scale that distinguishes the third family from the others, which doubles the number of mass terms for the scalar matter fields. If we assign a parameter $x_{LR}$ which quantifies the LR breaking in the scalar sector, one can derive the following relation for the SSB mass terms for the matter fields:
\begin{equation}
m_{\tilde{R}_{i}} = x_{LR}m_{\tilde{L}_{i}}~,~i=1,2 ~(3)~ {\rm for~ the~ first~ two~ (third)~ families}
\label{eq:PSgauginos}
\end{equation}

The relation among the SSB gaugino masses can be derived from the breaking of $422$. When the $422$ symmetry breaks to the MSSM gauge group the hypercharge generator remains unbroken and yields the following mass relation for the gauginos:
\begin{equation}
Y = \sqrt{\dfrac{3}{5}}I_{3R} + \sqrt{\dfrac{2}{5}}(B-L)\Rightarrow M_{1} = \dfrac{3}{5}M_{2R} + \dfrac{2}{5}M_{3}~,
\end{equation}
where $I_{3R}$ and $B-L$ are the diagonal generators of $SU(2)_{R}$ and $SU(4)_{c}$ respectively, and $M_{1}$, $M_{2R}$ and $M_{3}$ are the SSB mass terms for the gauginos associated with the $U(1)_{Y}$, $SU(2)_{R}$ and $SU(4)_{c}$ gauge groups respectively. As stated for the scalar matter fields, the LR symmetry breaking leads, in general, to $M_{2R}\neq M_{2L}$, where $M_{2L}$ denotes the mass of the $SU(2)_{L}$ gaugino. If we assign a parameter $y_{LR}$ to measure the LR breaking in the gaugino sector as $M_{2R}=y_{LR}M_{2L}$, Eq.(\ref{eq:PSgauginos}) yields
\begin{equation}
M_{1}=\dfrac{3}{5}y_{LR}M_{2L}+\dfrac{2}{5}M_{3}~.
\end{equation}

We can summarize the the set of GUT scale free parameters and their ranges in our scans as follows:
\begin{equation}
\begin{array}{lcl}
0 \leq & m_{\Tilde{L}_{1,2}}, m_{\Tilde{L}_{3}} & \leq 20 ~{\rm TeV} \\
0 \leq & M_{2L} & \leq 5 ~{\rm TeV} \\
-3 \leq & M_{3} & \leq 5 ~{\rm TeV} \\
-3 \leq & A_{0}/m_{\tilde{L}_{3}} & \leq 3 \\
35 \leq & \tan\beta & \leq 60 \\
0\leq & {\rm x}_{{\rm LR}} & \leq 3 \\
-3 \leq & {\rm y}_{{\rm LR}} & \leq 3 \\
0 \leq & {\rm x}_{{\rm d}} & \leq 3 \\
-1 \leq & {\rm x}_{{\rm u}} & \leq 2~, \\
\end{array}
\label{eq:paramSpacePSLR}
\end{equation}
where we assume a universal trilinear coupling denoted by $A_{0}$. Note that we consider the same $x_{LR}$ to quantify the LR symmetry breaking for all the families. We also impose non-universal Higgs boson masses at $\mgut$ which are parametrised as $m_{H_{d}}=x_{d}m_{\Tilde{L}_{3}}$ and $m_{H_{u}}=x_{u}m_{\Tilde{L}_{3}}$. In addition to the fundamental parameters, we only consider cases in which the bilinear Higgs mixing term is always positive ($\mu > 0$).

We perform random scans in the fundamental parameter space of $422$ by using SPheno \cite{Porod:2003um,Porod:2011nf} generated by SARAH \cite{Staub:2008uz,Staub:2015iza} for numerical calculations. After the GUT scale is determined through renormalization group equations (RGEs) by imposing the unification condition on the SM gauge couplings as $g_{1}=g_{2}\simeq g_{3}$, the RGEs runs back from $\mgut$ to $M_{Z}$ scale together with the SSB terms determined by the parameters given in Eq.(\ref{eq:paramSpacePSLR}). The SUSY mass spectra are calculated at the two-loop level. For the Higgs boson mass, SPheno employs also the self-energy contributions in addition to the two-loop corrections. Besides, we use the method which runs the SM RGEs at three-loop level between $M_{Z}$ and $\msusy$ by using the effective Higgs potential and imposing the matching condition at this scale. Note that we insert the central value of the top  quark mass ($m_{t}=173.3$ GeV \cite{CDF:2009pxd,ATLAS:2021urs}). Even though $1-2\sigma$ variations in the top quark mass do not affect the SUSY spectrum, it can yield a $1-2$ GeV shift in the SM-like Higgs boson mass \cite{AdeelAjaib:2013dnf,Gogoladze:2014hca}. In our scans, we accept only the solutions in which the LSP is one of the MSSM neutralinos. At the final step of the scans, we transfer the SPheno outputs to micrOMEGAs \cite{Belanger:2018ccd} to calculate the dark matter (DM) observables.  

In the scanning procedure, we employ the Metropolis-Hastings algorithm \cite{Baer:2008jn,Belanger:2009ti}. After generating the data, we successively employ the mass bounds \cite{Agashe:2014kda}, constraints from combined results for rare $B-$meson decays \cite{CMS:2020rox,Belle-II:2022hys,HFLAV:2022pwe}, and the latest Planck Satellite measurements \cite{Planck:2018nkj} of the DM relic abundance to constrain the LSP neutralino. We can summarize the experimental constraints employed in our analyses as follows: 
\begin{equation}
\setstretch{1.8}
\begin{array}{l}
m_h  = 123-127~{\rm GeV}\\
m_{\tilde{g}} \geq 2.1~{\rm TeV}~(800~{\rm GeV}~{\rm if~it~is~NLSP})\\
1.95\times 10^{-9} \leq{\rm BR}(B_s \rightarrow \mu^+ \mu^-) \leq 3.43 \times10^{-9} \;(2\sigma) \\
2.99 \times 10^{-4} \leq  {\rm BR}(B \rightarrow X_{s} \gamma)  \leq 3.87 \times 10^{-4} \; (2\sigma) \\
0.114 \leq \Omega_{{\rm CDM}}h^{2} \leq 0.126 \; (5\sigma)~.
\label{eq:constraints}
\end{array}
\end{equation} 
We have listed only the Higgs boson and gluino mass bounds, since they have been updated model independently, although we also employ the model independent mass bounds from the Linear electron-positron collider (LEP2) \cite{ParticleDataGroup:2014cgo}. Even though the parameters listed in Eq.(\ref{eq:constraints}) have been measured experimentally with significant precision, we consider a few $\sigma$ variations to compensate for the uncertainties in their theoretical calculations arising from strong interaction coupling, the top quark mass, the mixing in the squark sectors, etc. \cite{Gogoladze:2011aa,AdeelAjaib:2013dnf,Degrassi:2002fi,Bahl:2019hmm,Bagnaschi:2017xid, Athron:2016fuq,Allanach:2004rh,Drechsel:2016htw}. We employ a $5\sigma$ uncertainty in constraining the relic abundance of LSP neutralino since the uncertainties in its theoretical calculations exceed the statistical uncertainties in its experimental measurements \cite{Bergeron:2017rdm,Baer:2021tta}. We identify the solutions compatible with the QYU condition if the deviation in Yukawa couplings quantified by the $C$ parameter satisfies $|C| \leq 0.2$. Note that Eq.(\ref{eq:CQYU}) leads to three different solutions for $C$ in terms of different combinations of the Yukawa couplings. We also require the QYU compatible solutions to yield the same $|C|$ in all three Yukawa couplings up to about $10\%$ uncertainty.

\section{QYU and Mass Spectrum}
\label{sec:fundQYU}

\begin{figure}[t!]
\centering
\subfigure{\includegraphics[scale=0.4]{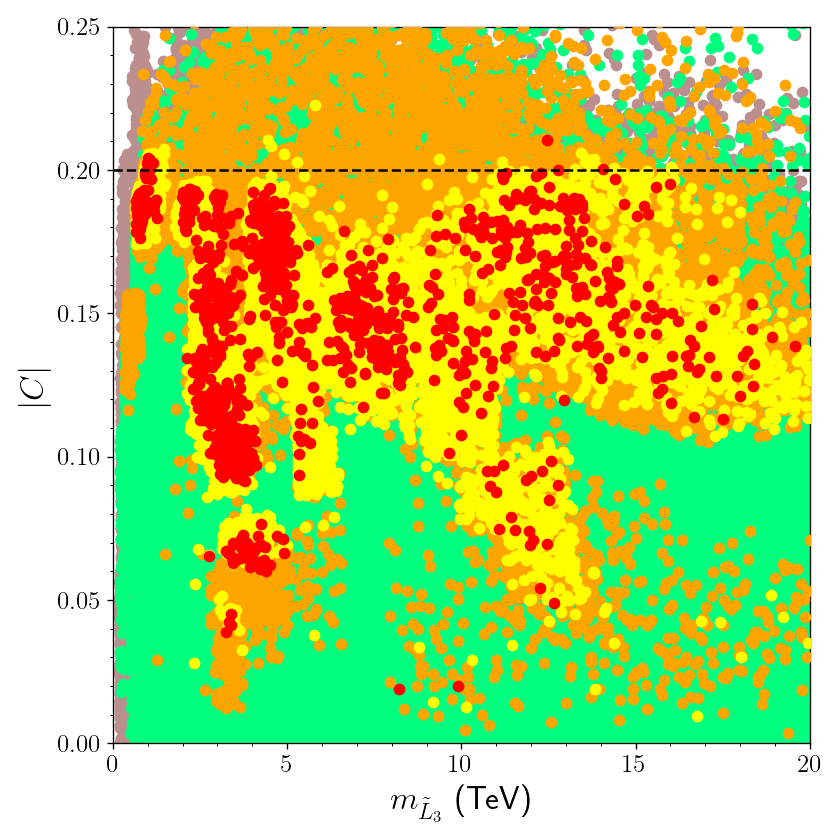}}%
\subfigure{\includegraphics[scale=0.4]{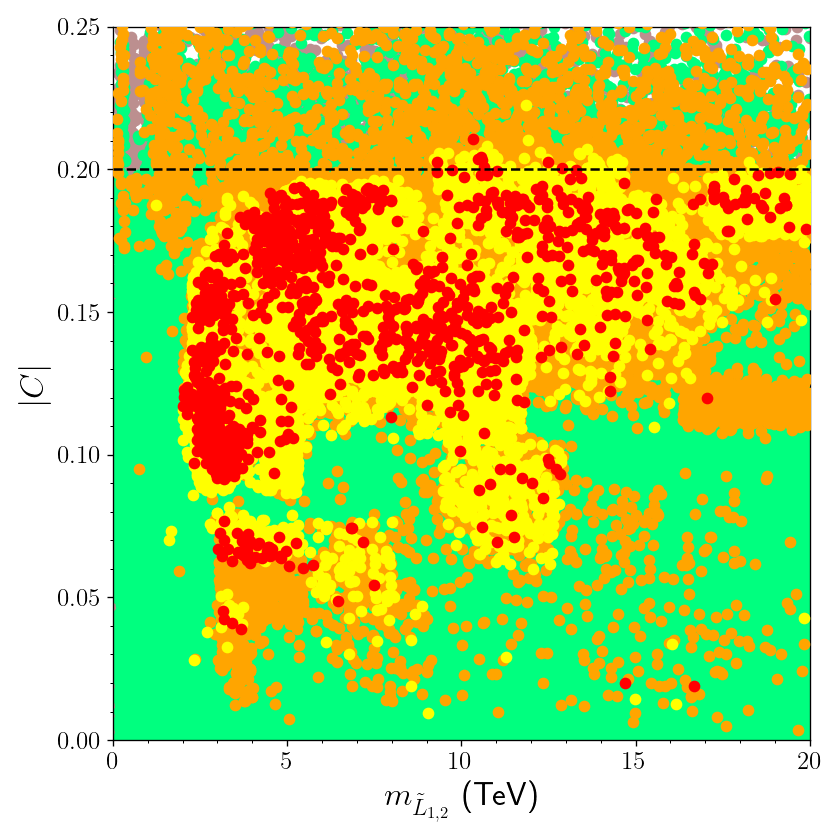}}
\subfigure{\includegraphics[scale=0.4]{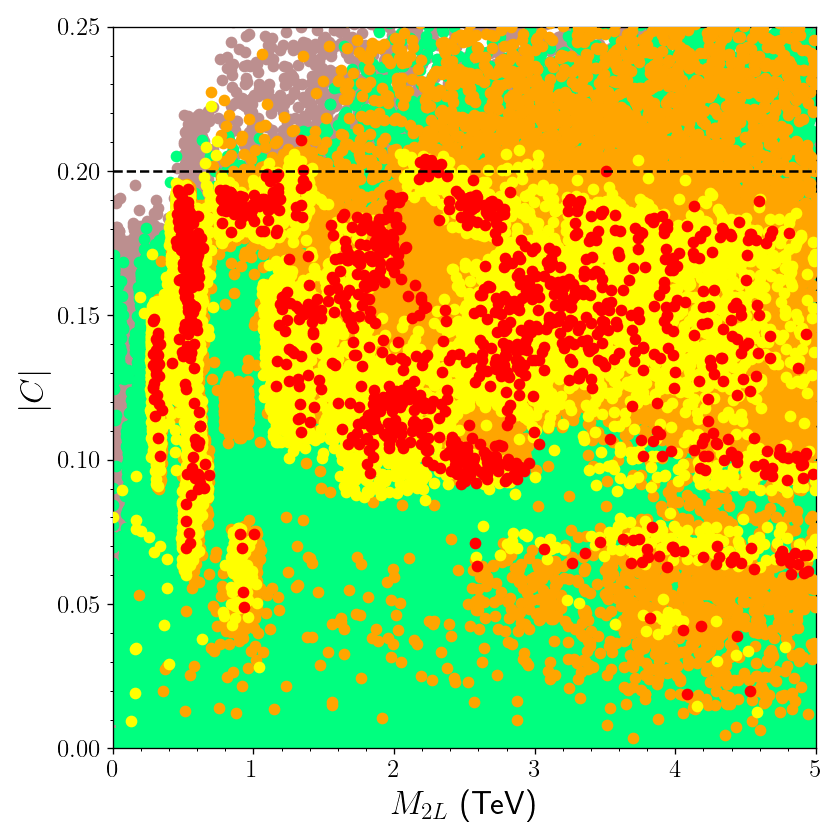}}%
\subfigure{\includegraphics[scale=0.4]{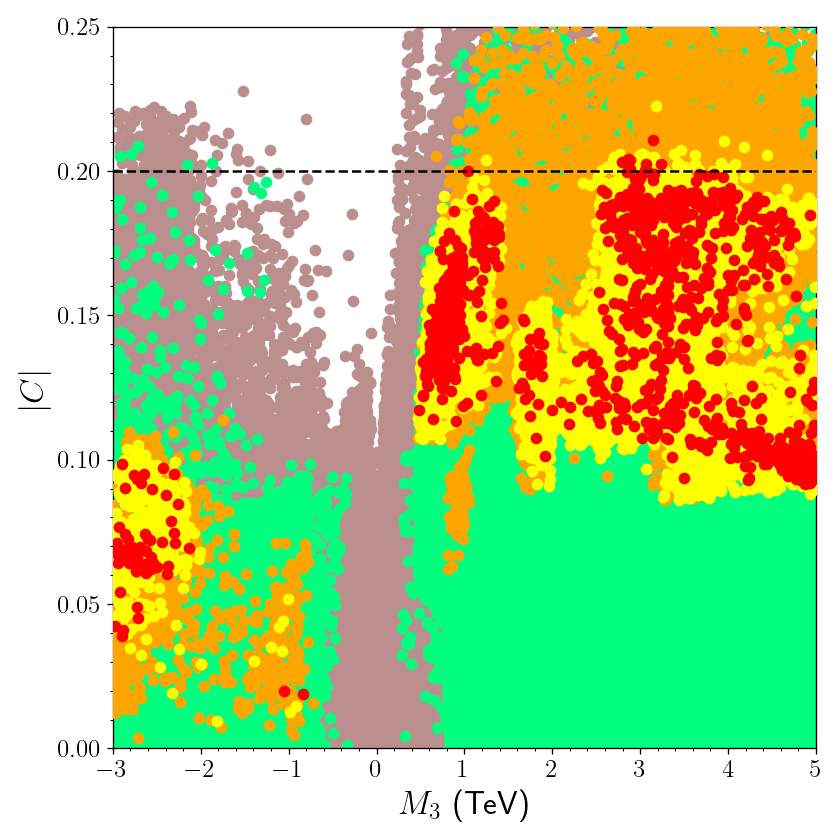}}
\caption{Plots in the $C-m_{\tilde{L}_{3}}$, $C-m_{\tilde{L}_{1,2}}$, $C-M_{2L}$ and $C-M_{3}$ planes. All solutions are compatible with REWSB and LSP neutralino conditions. Green points satisfy the mass bounds on the sparticles and the constraints from rare B-decays. Orange points form a subset of green and they are compatible with the QYU condition. Red points are a subset of orange and satisfy the constraint on relic abundance of LSP neutralino from Planck measurements within $5\sigma$. Yellow points form a subset of orange with DM relic density lower than the $5\sigma$ constraints from the Planck measurements.}
\label{fig:GUTmasses}
\end{figure}

\begin{figure}[t!]
\centering
\subfigure{\includegraphics[scale=0.4]{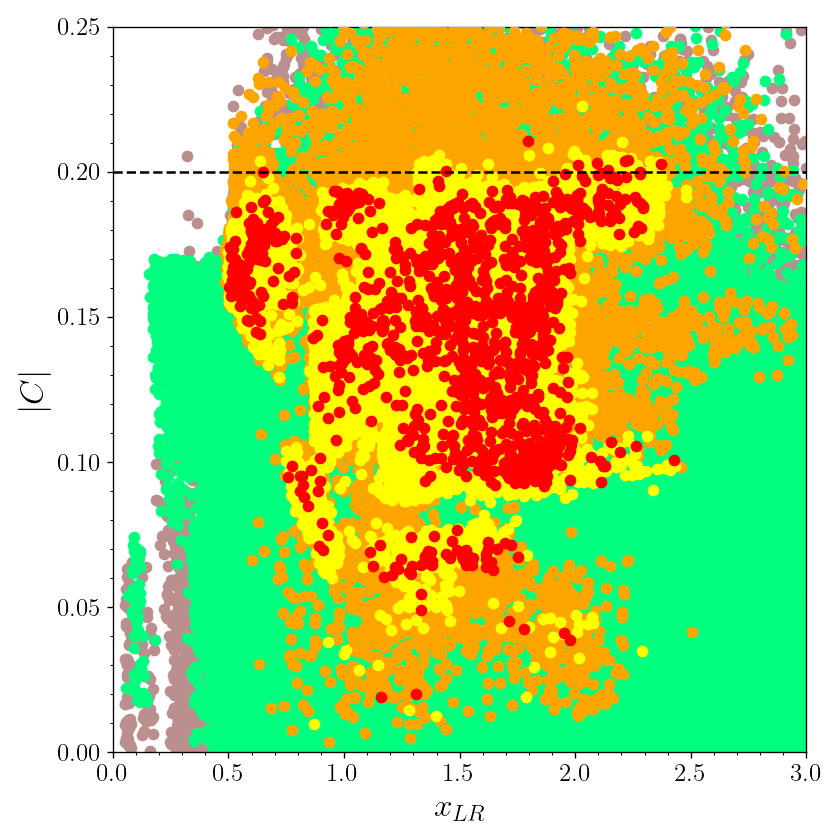}}%
\subfigure{\includegraphics[scale=0.4]{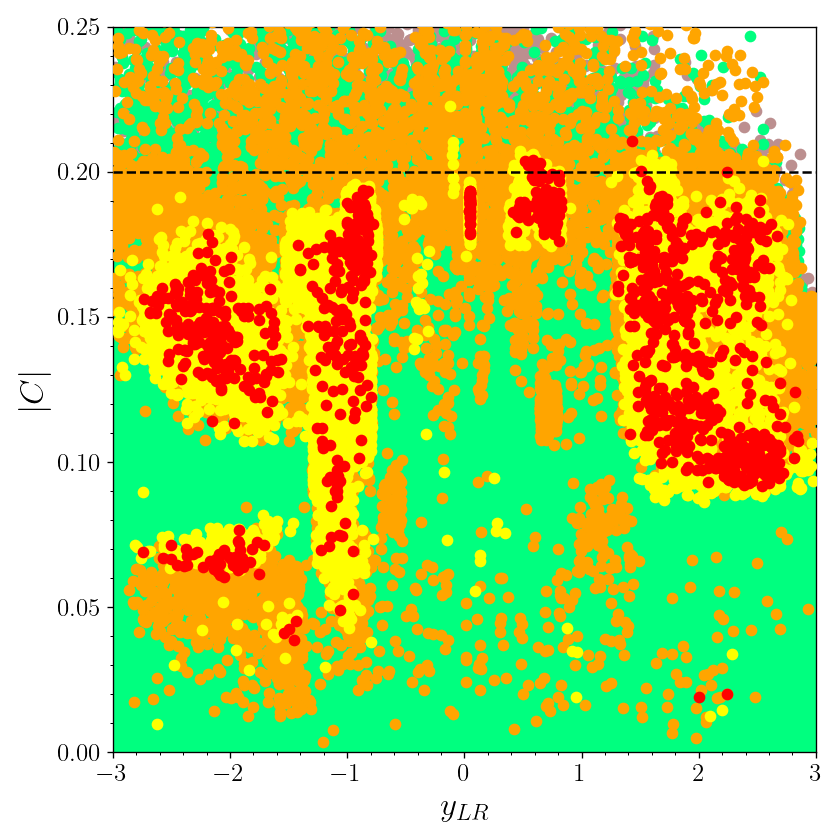}}
\caption{Plots in the $C-x_{LR}$ and $C-y_{LR}$ planes. Color coding is the same as in Figure \ref{fig:GUTmasses}.}
\label{fig:LRbreaking}
\end{figure}

\begin{figure}[t!]
\centering
\subfigure{\includegraphics[scale=0.4]{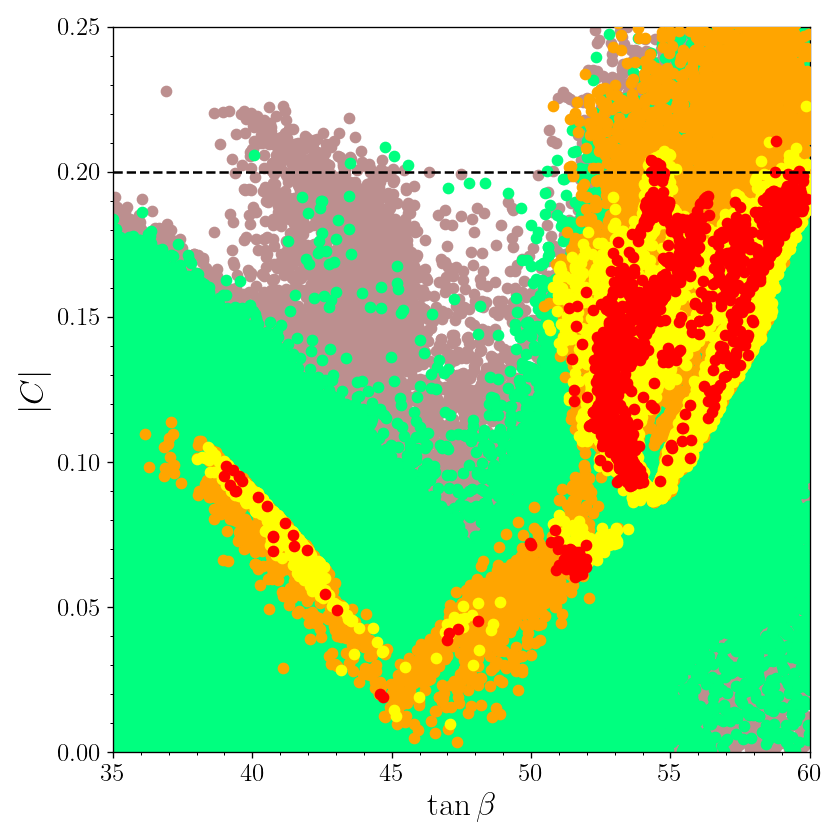}}%
\subfigure{\includegraphics[scale=0.4]{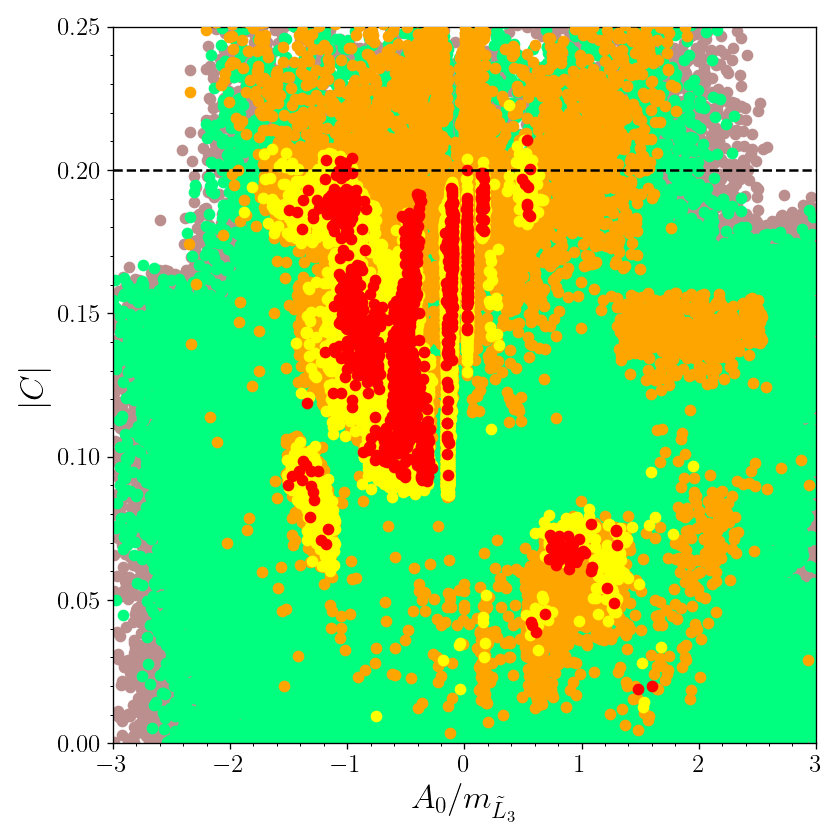}}
\subfigure{\hspace{0.6cm}\includegraphics[width=7.0cm,height=7.0cm]{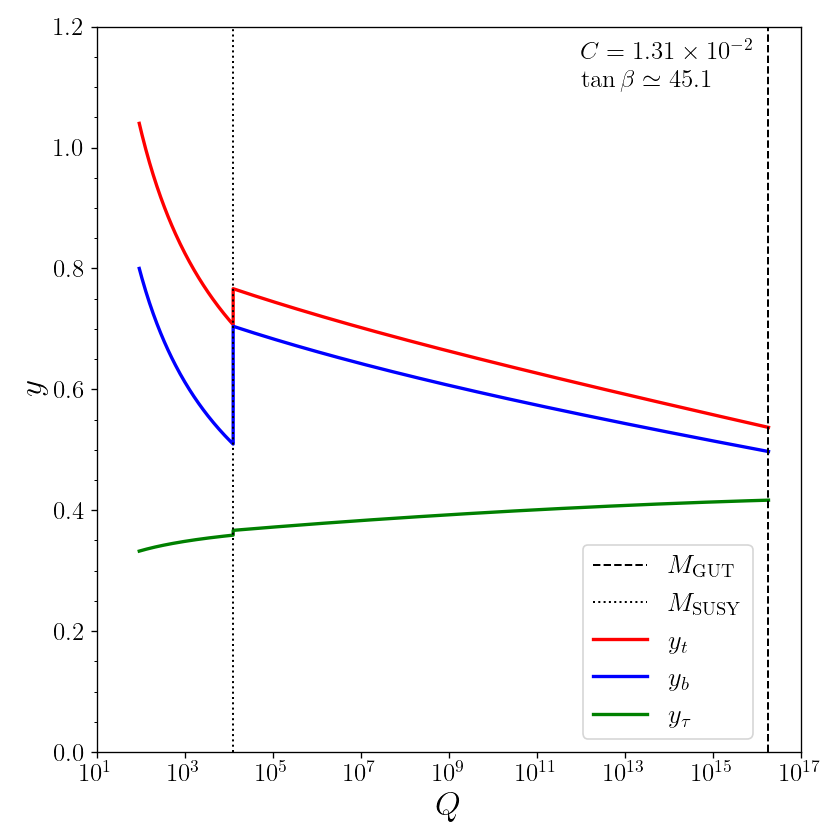}}%
\subfigure{\hspace{0.1cm}\includegraphics[width=7.0cm,height=7.0cm]{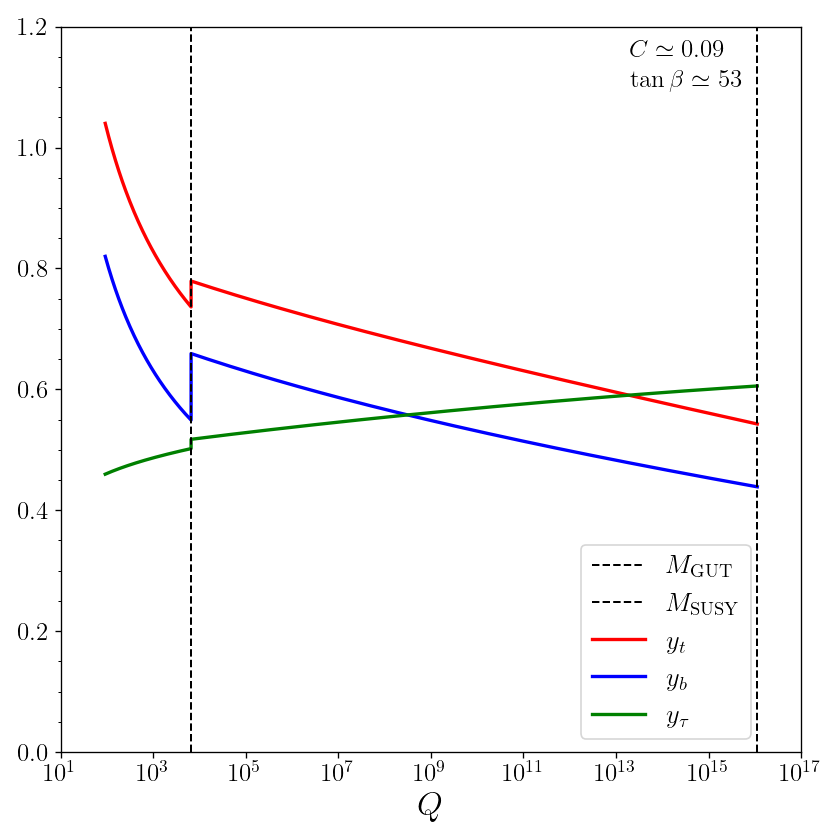}}
\caption{Plots in the $C-A_{0}/m_{3\Tilde{L}}$ and $C-\tan\beta$ planes (top), and the RGE evolution of the Yukawa couplings (bottom). Color coding in the top planes is the same as in Figure \ref{fig:GUTmasses}. The curves and vertical lines in the bottom planes are defined in the panels.}
\label{fig:AtanRGEs}
\end{figure}

\begin{figure}[t!]
\centering
\subfigure{\includegraphics[scale=0.4]{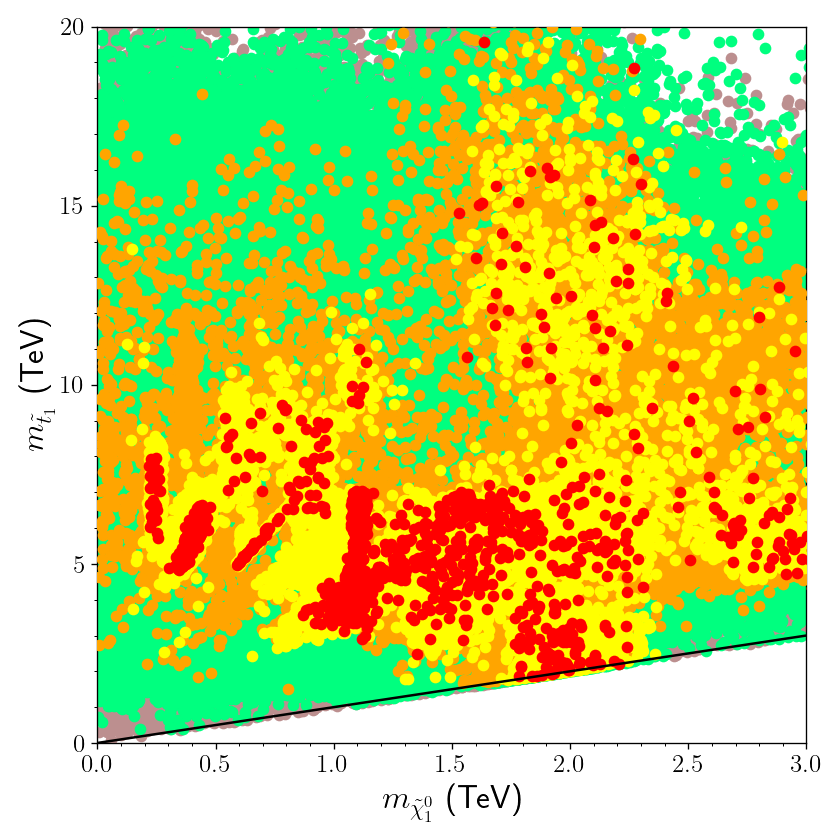}}%
\subfigure{\includegraphics[scale=0.4]{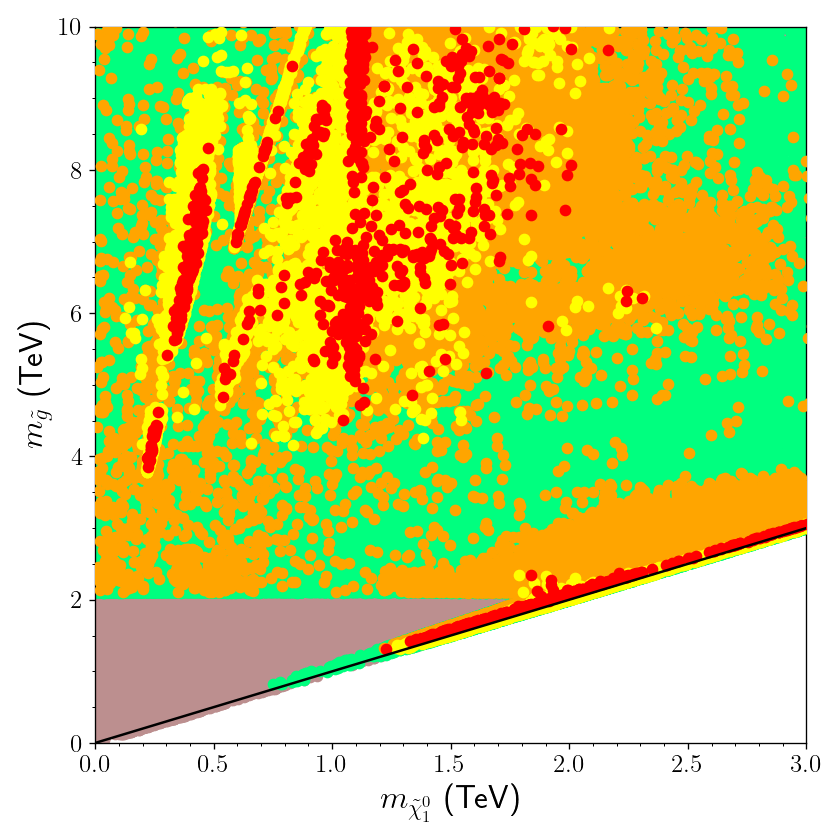}}
\caption{Plots in the $m_{\tilde{t}_{1}}-m_{\tilde{\chi}_{1}^{0}}$ and $m_{\tilde{g}}-m_{\tilde{\chi}_{1}^{0}}$ planes. All solutions are compatible with REWSB and LSP neutralino conditions. Green points satisfy the mass bounds on the sparticles and the constraints from rare B-decays. Orange points form a subset of green and satisfy $\mid C \mid \leq 0.2$ as well as the QYU condition. Points in red form a subset of orange and satisfy the constraint on relic abundance of LSP neutralino from Planck measurements within $5\sigma$. Points in yellow form a subset of orange with lower relic density than the Planck measurements. The diagonal lines represent solutions in which the particles shown are degenerate in mass.}
\label{fig:stopglu}
\end{figure}

In this section, we display the plots for the GUT scale mass parameters in Figure \ref{fig:GUTmasses} with plots in the $C-m_{\tilde{L}_{3}}$, $C-m_{\tilde{L}_{1,2}}$, $C-M_{2L}$ and $C-M_{3}$ planes. All solutions are compatible with REWSB and LSP neutralino conditions. Green points satisfy the mass bounds on the sparticles and the constraints from rare B-decays. Orange points form a subset of green and they are compatible with the QYU condition. Red points are a subset of orange and satisfy the constraint on relic abundance of LSP neutralino from Planck measurements within $5\sigma$. Yellow points form a subset of orange with DM relic density lower than the $5\sigma$ constraints from the Planck measurements. The $C-m_{\tilde{L}_{3}}$ plane shows that the QYU condition on the Yukawa couplings at $\mgut$ can be realized in a wide range as $0.3 \lesssim \mLT \lesssim 20$ TeV. The solutions with low $\mLT$ yield a relatively large deviation from YU ($|C| \gtrsim 0.12$), and the DM relic density bounds it at about 0.18 from below (red solutions). Besides, the lighter solutions lead to the SM-like Higgs boson mass at about 123 GeV, and the central value of the Higgs boson mass ($\sim 125.6$ GeV) favors the solutions with $m_{\tilde{L}_{3}} \gtrsim 1$ TeV. The SSB mass term for the first two families compatible with the QYU condition can also be light, but the solutions consistent with the DM relic density can be realized if $\mLFS \gtrsim 3$ TeV, as shown in the $|C|-\mLFS$ plane. The bottom panels of Figure \ref{fig:GUTmasses}. Similar observation can be seen from the bottom panels where the SSB masses for $SU(2)_{L}$ and $SU(3)_{c}$ gauginos are plotted. Note that the fade region in the $C-M_{3}$ plane ($-600 \lesssim M_{3}\lesssim 600$ GeV) is excluded by the current mass bound on gluino.

The masses for the right-handed partners of the SUSY scalars and gauginos can be considered through the LR symmetry breaking parametrized by $x_{LR}$ and $y_{LR}$ which are displayed in Figure \ref{fig:LRbreaking} with plots in the $C-x_{LR}$ and $C-y_{LR}$ planes. Color coding is the same as in Figure \ref{fig:GUTmasses}. The $C-x_{LR}$ plane shows that LR breaking in the scalar sector is compatible with the QYU condition and the other constraints can measure in the range $0.7 \lesssim x_{LR} \lesssim 2.2$. Even though it is possible to realize QYU if the scalar sector is LR symmetric ($x_{LR}=1$), as seen from the $C-y_{LR}$ plane, QYU is mostly realized if LR symmetry is broken in the gaugino sector ($|y_{LR}| \gtrsim 1$). In addition, large LR breaking in the gaugino sector can allow very small deviations in YU ($|C| \gtrsim 2\%$). Note that such solutions are realized also for $y_{LR} \sim 1$ and $|C| \sim (1-1.2)\times 10^{-2}$, which can be considered recovering YU. However, these solutions yield a large relic abundance of LSP neutralino and are excluded by the current Planck measurements within $5\sigma$.

We continue our discussion with the correlation of QYU with $\tan\beta$ and the SSB trilinear scalar interacting term in Figure \ref{fig:AtanRGEs}, which also shows the RGE evolution of the Yukawa couplings between $\mgut$ and $M_{Z}$. Color coding in the top planes is the same as in Figure \ref{fig:GUTmasses}. The curves and vertical lines in the bottom planes are defined in the legend. The $C-\tan\beta$ plane shows a nearly linear correlation between $C$ and $\tan\beta$, which indicates that the deviation from YU increases proportionally with $\tan\beta$. One can realize a negligible deviation for $\tan\beta \sim 45$. This value of $\tan\beta$ is also suitable for exact YU (see, for instance, \cite{Hussain:2018xiy,Antusch:2019gmc,Gomez:2020gav,Hicyilmaz:2021onw}). The results in the $C-A_{0}/m_{3\tilde{L}}$ plane do not show a specific correlation between $C$ and the trilinear scalar interacting term, and the QYU condition requires $-2 \lesssim A_{0}/m_{3\tilde{L}} \lesssim 3$.

These relations, especially one involving $\tan\beta$, can be understood through the RGE evolution of the Yukawa couplings shown in the bottom plane of Figure \ref{fig:AtanRGEs} for different $\tan\beta$ values. The Yukawa couplings receive threshold correction at $\msusy \equiv \sqrt{m_{\tilde{t}_{R}}m_{\tilde{t}_{L}}}$, below which the SUSY particles are assumed to decouple, where $m_{\tilde{t}_{L,R}}$ stand for the left-handed and right-handed stops. These threshold corrections play an important role in realizing YU consistent with the observed third-family fermion masses. The bottom-quark Yukawa coupling, in particular, needs large and negative threshold corrections in the case of YU \cite{Gogoladze:2010fu}, which can be recovered in our model for $\tan\beta \sim 45$ and $|C| \sim 0$. These cases are represented with the point whose RGE evolution is shown in the bottom-left plane of Figure \ref{fig:AtanRGEs}. We note that $y_{b}$ needs a negative threshold correction as large as $|\delta y_{b}| \simeq 0.2$ to yield a consistent bottom quark mass at $M_{Z}$. However, moving away from the YU region, the impact of the threshold corrections can be loosened. The bottom-right panel exemplifies such solutions with correct fermion masses realized even with $|\delta y_{b}| \sim 0.1 $. Note that the curves shown crossing in the bottom figures do not indicate unification at low scales, but occur only numerically in their RGE evolution.

\begin{equation}
\delta y_{b} \approx \dfrac{g_{3}^{2}}{12\pi^{2}}\dfrac{\mu m_{\tilde{g}}\tan\beta}{m_{\tilde{b}}^{2}} + \dfrac{y_{t}^{2}}{32\pi^{2}}\dfrac{\mu A_{t}\tan\beta}{m_{\tilde{t}}^{2}}.
\label{eq:deltayb}
\end{equation}

Eq.(\ref{eq:deltayb}) also reveals the impact of the threshold corrections on the parameters and mass spectrum. For $\mu > 0$, the contributions from the gluino loop should be suppressed which leads to next to LSP (NLSP) gluino solutions of mass $m_{\tilde{g}} \lesssim 1$ TeV \cite{Raza:2014upa,Gogoladze:2009ug,Gogoladze:2009bn}. It also explains why one needs an appreciable LR breaking ($|y_{LR}| \sim 3$ as shown in Figure (\ref{fig:LRbreaking}) to realize small deviations in YU ($|C| \sim 0$). However, as discussed above, the QYU solutions do not necessarily employ large threshold corrections, so the upper bound on the gluino mass from the YU condition disappears in our analyses. We display the results for the stop and gluino masses versus the LSP neutralino mass in Figure \ref{fig:stopglu} in the $m_{\tilde{t}_{1}}-m_{\tilde{\chi}_{1}^{0}}$ and $m_{\tilde{g}}-m_{\tilde{\chi}_{1}^{0}}$ planes. Color coding is the same as in Figure \ref{fig:GUTmasses}. The diagonal lines designate the solutions with degenerate masses. We obtain mass spectra in which the stop mass cannot be lighter than about 1.5 TeV in the QYU parameter space (orange points), while the relic density constraint raises this lower mass bound to about 1.8 TeV (red points), where it also happens to be degenerate with the LSP neutralino in mass. These nearly degenerate solutions play an important role in reducing the relic abundance of LSP neutralino through stop-neutralino coannihilation scenario. The heavy mass scales for the stop are mostly required in order to obtain the correct SM Higgs boson mass \cite{Gomez:2022qrb}. On the other hand, we realize NLSP gluino solutions compatible with QYU condition with $m_{\tilde{g}}\gtrsim 800$ GeV, and the DM relic density constraint can be satisfied for $m_{\tilde{g}}\gtrsim 1$ TeV. The possibility of small threshold corrections also allows heavy gluino masses, and our scans yield gluino masses up to about 10 TeV. Although testing such heavy gluino solutions requires a much higher center of mass energies and luminosities than currently available at the LHC, the solutions with $2.1 \leq m_{\tilde{g}} \lesssim 2.5$ TeV should be testable during the LHC-Run3 experiments \cite{ATLAS:2022rcw}.

\section{DM Implications of QYU}
\label{sec:DM}

\begin{figure}[t!]
\centering
\subfigure{\includegraphics[scale=0.4]{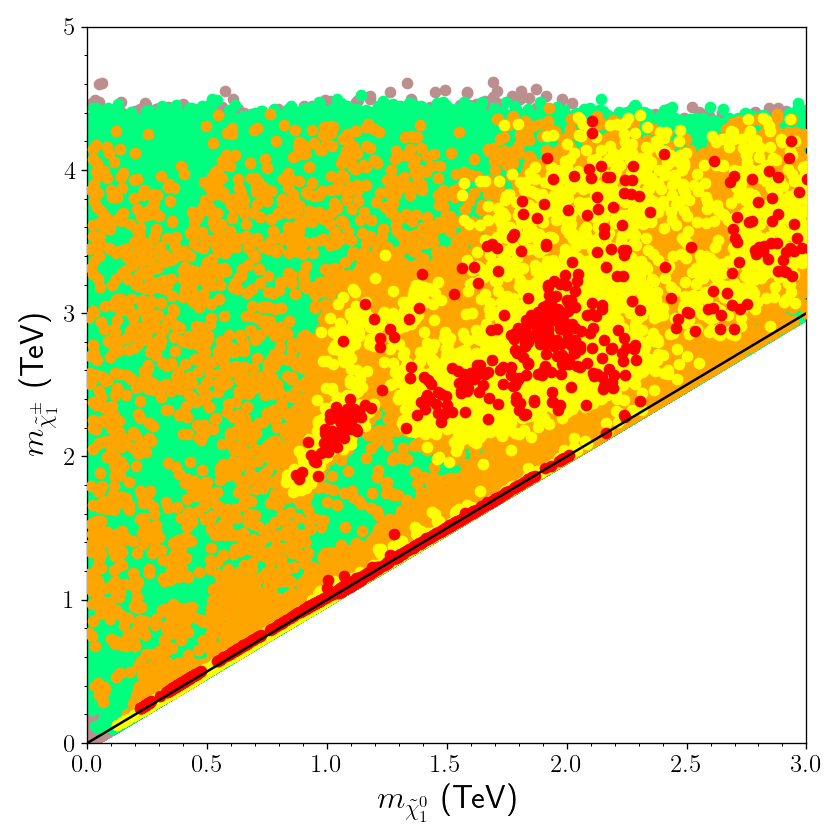}}%
\subfigure{\includegraphics[scale=0.4]{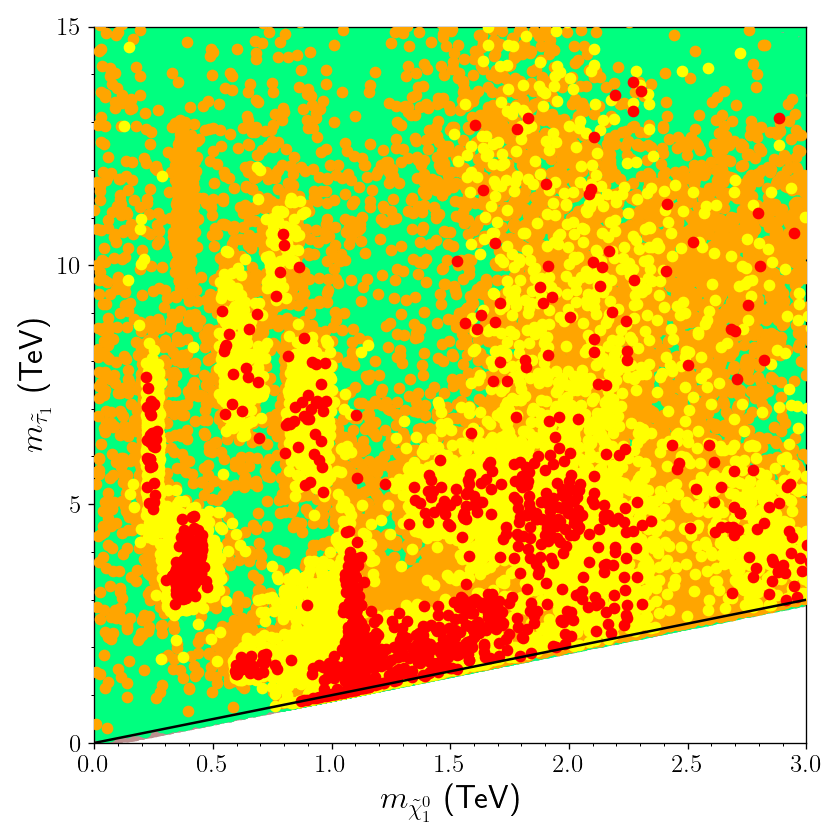}}
\subfigure{\includegraphics[scale=0.4]{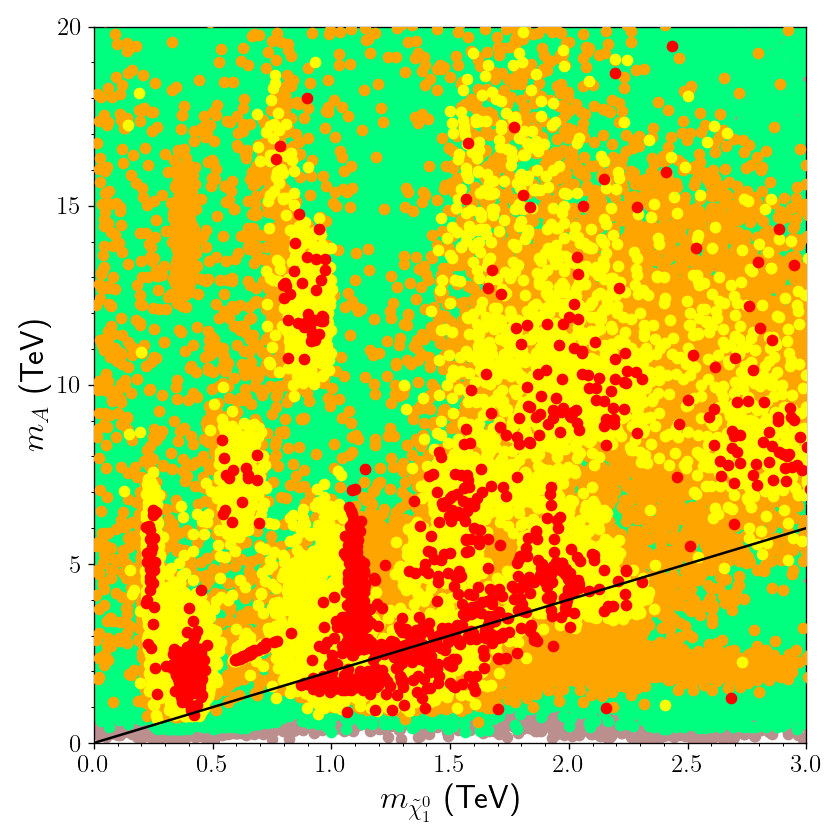}}%
\subfigure{\includegraphics[scale=0.4]{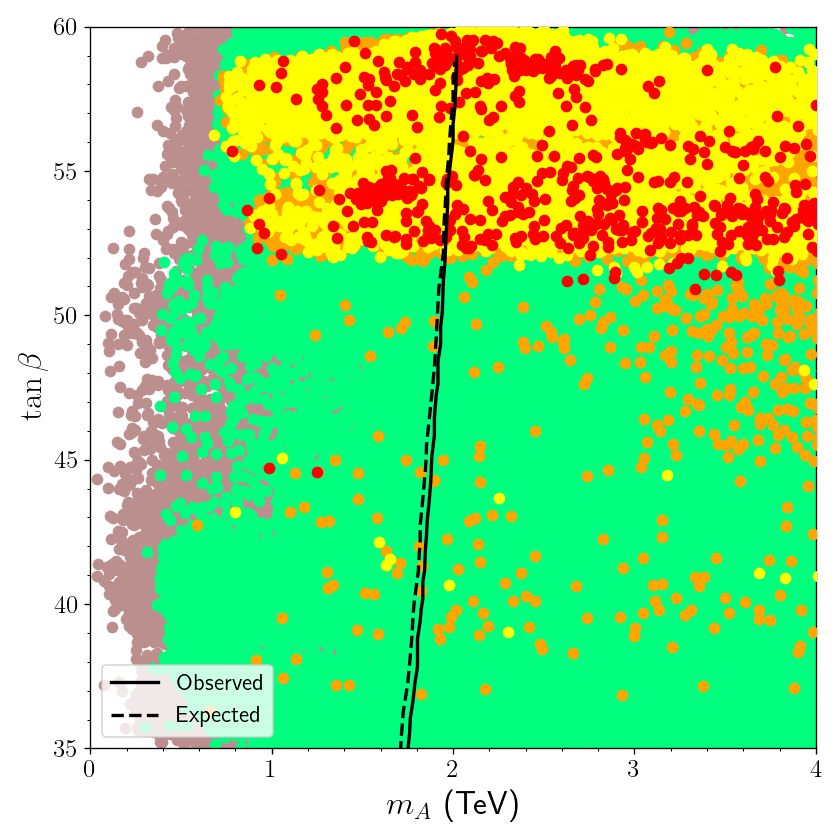}}
\caption{Plots in the $m_{\tilde{\chi}_{1}^{\pm}}-m_{\tilde{\chi}_{1}^{0}}$, $m_{\tilde{\tau}_{1}}-m_{\tilde{\chi}_{1}^{0}}$, $m_{A}-m_{\tilde{\chi}_{1}^{0}}$ and $\tan\beta - m_{A}$ planes. The color coding is the same as in Figure \ref{fig:stopglu}. The diagonal lines in the top panels indicate the mass degenerate solutions for the particles shown, and the line in the $m_{A}-m_{\tilde{\chi}_{1}^{0}}$ plane represents $A-$resonance solutions ($m_{A}=2m_{\tilde{\chi}_{1}^{0}}$). The curves in the $\tan\beta - m_{A}$ plane display the current exclusion region \cite{Bagnaschi:2018ofa,ATLAS:2020zms} for the CP-odd Higgs boson mass.}
\label{fig:coans}
\end{figure}

\begin{figure}[t!]
\centering
\subfigure{\includegraphics[scale=0.4]{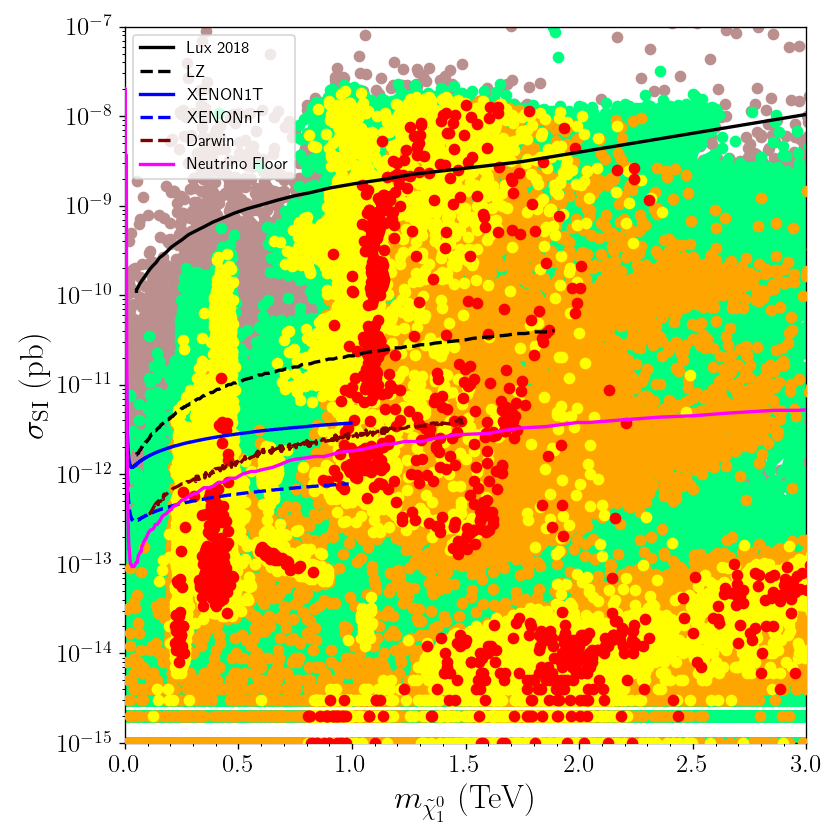}}%
\subfigure{\includegraphics[scale=0.4]{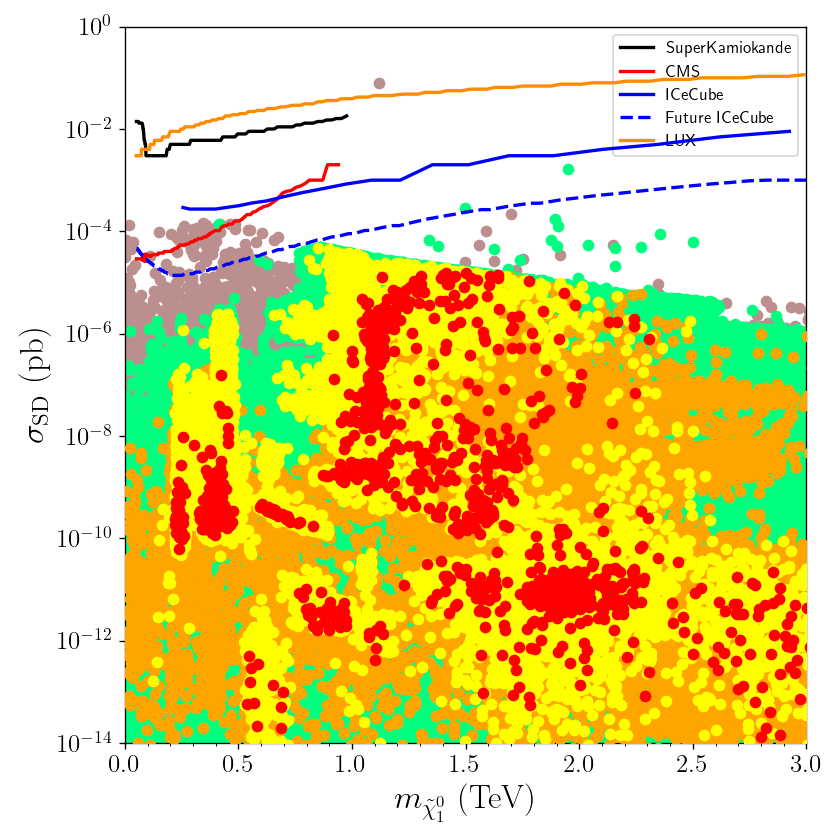}}
\caption{Spin-independent (left) and spin-dependent (right) scattering cross-sections versus the LSP neutralino mass. The color coding is the same as in Figure \ref{fig:stopglu}. The curves represent the current and projected exclusion curves from several direct detection DM experiments, with the color coding given in the respective panels. The current excluded regions are represented by the solid curves, and the dashed curves display the projected experimental sensitivity.}
\label{fig:SISD}
\end{figure}

Even though the MSSM provides a variety of different neutralino species (Bino, Wino, Higgsinos) as DM candidates, the current relic density and direct detection experiments have yielded a strong impact on the DM implications. The Higgsino-like LSPs usually lead to large cross-sections in scattering with nuclei, and together with the relic density constraints the current results exclude the Higgsino-like LSP solutions if they are lighter than about 1 TeV (see, for instance, \cite{Raza:2018jnh,Hicyilmaz:2021khy}). Even though their scattering cross-section is moderate, the Wino-like LSP solutions are similarly constrained by the experiments. On the other hand, Bino-like LSP has a relatively small scattering cross-section. However, considering the current and future projected sensitivities in the direct detection experiments \cite{Akerib:2018lyp,Aalbers:2016jon,Aprile:2020vtw,Tanaka:2011uf,Khachatryan:2014rra,Abbasi:2009uz,Akerib:2016lao}, these cross-sections are expected to be tested soon.

The desired DM relic density with Bino-like LSP requires suitable coannihilation processes, and as shown in Section \ref{sec:fundQYU}, this is satisfied for a gluino NLSP mass $m_{\tilde{g}} \gtrsim 1.3$ TeV. In addition, the stop-neutralino coannihilation scenario is realized for $1.8 \lesssim m_{\tilde{t}_{1}}\simeq m_{\tilde{\chi}_{1}^{0}} \lesssim 2.5$ TeV.

In addition, we also identify chargino-neutralino and stau-neutralino coannihilations scenarios, as well as $A-$resonance solutions as shown in Figure \ref{fig:coans}, with plots in the $m_{\tilde{\chi}_{1}^{\pm}}-m_{\tilde{\chi}_{1}^{0}}$, $m_{\tilde{\tau}_{1}}-m_{\tilde{\chi}_{1}^{0}}$, $m_{A}-m_{\tilde{\chi}_{1}^{0}}$ and $\tan\beta - m_{A}$ planes. The color coding is the same as in Figure \ref{fig:GUTmasses}. The diagonal lines in the top panels indicate the mass degenerate solutions for the particles shown, and the $m_{A}-m_{\tilde{\chi}_{1}^{0}}$ plane displays $A-$resonance solutions ($m_{A}=2m_{\tilde{\chi}_{1}^{0}}$). The curves in the $\tan\beta - m_{A}$ plane display the current exclusion limits \cite{Bagnaschi:2018ofa,ATLAS:2020zms} on the CP-odd Higgs boson mass versus $\tan\beta$. The $m_{\tilde{\chi}_{1}^{\pm}}-m_{\tilde{\chi}_{1}^{0}}$ plane shows viable chargino-neutralino coannihilation solutions compatible with the DM abundance in a wide range with the chargino mass varying from about 250 GeV to 2.5 TeV. The solutions with $m_{\tilde{\chi}_{1}^{0}} \lesssim 1.5$ TeV around the diagonal line with the correct relic density (red points) usually lead to Bino-like relic density with $M_{\tilde{B}} \lesssim M_{\tilde{W}}$, since the Wino-like and Higgsino-Like LSP solutions ($M_{\tilde{W}} \ll M_{\tilde{B}}$) are mostly excluded by the relic density constraints for $m_{\tilde{\chi}_{1}^{0}} \lesssim 1.5$ TeV (see, for instance, \cite{Frank:2021nkq}). We also display solutions with stau-neutralino coannihilations in the $m_{\tilde{\tau}_{1}}-m_{\tilde{\chi}_{1}^{0}}$ plane. Even though the stau is heavy over most of the parameter space due to the QYU condition, the stau-neutralino coannihilation solutions can be realized if the stau and LSP neutralino are nearly degenerate in the mass range of about $1-2.5$ TeV.

Besides the coannihilation channels, the QYU scenario admits self annihilations of LSP neutralinos into the CP-odd Higgs boson in a wide range, namely $0.5 \lesssim m_{A} \lesssim 2.5$ TeV, as shown in the $m_{A}-m_{\tilde{\chi}_{1}^{0}}$ plane. The mass scale for $m_{A}$ is constrained by the $A\longrightarrow\tau \tau$ decay, which is displayed with the dotted and solid curves in the $\tan\beta - m_{A}$ plane. With a mass heavier than a few hundred GeV, the neutral heavy Higgs bosons ($A$ and $H$) decay into a pair of $\tau-$leptons at a level of about $15\%$, which is large enough to exclude the solutions with $m_{A}\lesssim 2$ TeV \cite{Bagnaschi:2018ofa,ATLAS:2020zms}. A recent analysis \cite{CMS:2018oxh} has shown that experiments at the LHC-Run3 and High-Luminosity LHC (HL-LHC) can probe the CP-odd Higgs boson up to about 2.5 TeV. Our results in the $\tan\beta - m_{A}$ plane show that QYU solutions are abundantly realized for $2 < m_{A} < 2.5$ TeV, which should be tested at LHC Run-3.

Another probe for the QYU solutions can be provided in the direct detection experiments of DM. Figure \ref{fig:SISD} shows our results for the cross-sections for the spin-independent (left) and spin-dependent (right) scattering of DM on nuclei versus the LSP mass. The color coding is the same as in Figure \ref{fig:GUTmasses}. The solid (dashed) curves represent the current (future projected) exclusion curves from several direct detection DM experiments, with color coding given in the legend for each plane. The $\sigma_{{\rm SI}}-m_{\tilde{\chi}_{1}^{0}}$ shows that the LZ experiment \cite{Akerib:2018lyp} can currently exclude the QYU solutions with spin-independent cross-section on the order of about $4\times 10^{-11}$ pb and the LSP of about 1 TeV, while XENON1T has extended this exclusion limit to about $4\times 10^{-12}$ pb in the same region. The solutions whose cross-sections are above these bounds will be re-tested by the Darwin experiment in the near future \cite{Aalbers:2016jon}. Furthermore, the projected sensitivity of XENON experiment (the curve from XENONnT) should be able to probe QYU solutions at the level of about $8\times 10^{-13}$ pb \cite{Aprile:2020vtw}. We also present our results for the spin-dependent scattering cross-section, and as seen from the $\sigma_{{\rm SD}}-m_{\tilde{\chi}_{1}^{0}}$, with the cross-sections a few orders of magnitude lower than the experimental bounds \cite{Tanaka:2011uf,Khachatryan:2014rra,Abbasi:2009uz,Akerib:2016lao}, we have to wait for further upgrades in the experiments listed in the legend.

\begin{table}[t!]
\centering 
\scalebox{0.9}{
\begin{tabular}{|c|cccccc|}
\hline  & Point 1 & Point 2 & Point 3 & Point 4 & Point 5 & Point 6 \\ \hline 
$m_{\tilde{L}_{1,2}}$ & 5016 & 9007 & 10049 & 3037 & 5063 & 2981 \\
$m_{\tilde{L}_{3}}$ & 5316 & 7167 & 8031 & 2493 & 4562 & 3153 \\
$M_{1}$ & 998.7 & -2638 & -3842 & 2544 & \textbf{3672} & 7675 \\
$M_{2}$ & 585.1 & 2803 & 2853 & 1390 & \textbf{1989} & 4797 \\
$M_{3}$ & 3564 & 484.3 & 727.4 & 2990 & 4313 & 4230 \\
$A_{0}/m_{\tilde{L}_{3}}$ & -0.191 & -1.5 & -2 & -0.793 & -0.76 & -1.1 \\
$\tan\beta$ & 55.4 & 52.2 & 54.1 & 54.8 & 55.8 & 53.1 \\ \hline 
$x_{LR}$ & 1.3 & 1.6 & 1.8 & 1.6 & 1.7 & 2.1 \\
$y_{LR}$ & -1.2 & -1.7 & -2.4 & 1.6 & 1.6 & 2.1 \\
$m_{\tilde{R}_{1,2}}$ & 6405 & 14051 & 18517 & 4782 & 8748 & 6298 \\
$m_{\tilde{R}_{3}}$ & 6788 & 11181 & 14799 & 3926 & 7884 & 6662 \\ \hline 
$\mu$ & 1627 & 8762 & 10877 & 2807 & 5327 & \textbf{1086} \\
$m_{h}$ & 125 & 124 & 125.7 & 123.9 & 125.3 & 125.6 \\
$m_{H}$ & 2973 & 4973 & 4660 & 3578 & 3878 & 2447 \\
$m_{A}$ & 2973 & 4973 & 4660 & 3578 & 3878 & \textbf{2447} \\
$m_{H^{\pm}}$ & 2974 & 4972 & 4655 & 3579 & 3877 & 2450 \\ \hline 
$m_{\tilde{\chi}_{1}^{0}}$,$m_{\tilde{\chi}_{2}^{0}}$ & \red{426.6}, \red{454.7} & \red{1227}, 2466 & \red{1785}, 2496 & \red{1137}, \red{1153} & \red{1674}, \red{1676} & \red{1087}, \red{1090} \\
$m_{\tilde{\chi}_{3}^{0}}$,$m_{\tilde{\chi}_{4}^{0}}$ & 1632, 1634 & 8691, 8691 & 10915, 10915 & 2815, 2817 & 5326, 5326 & 3520, 3992 \\
$m_{\tilde{\chi}_{1}^{\pm}}$,$m_{\tilde{\chi}_{2}^{\pm}}$ & \red{454.8}, 1634 & 2466, 8691 & 2496, 10915 & \red{1153}, 2817 & \red{1677}, 5327 & \red{1088}, 3991 \\ \hline 
$m_{\tilde{g}}$ & 7509 & \red{1319} & 1914 & 6258 & 8920 & 8621 \\
$m_{\tilde{u}_{1}}$,$m_{\tilde{u}_{2}}$ & 7854, 8783 & 9098, 14023 & 10080, 18539 & 6031, 7012 & 8835, 11306 & 8184, 9612 \\
$m_{\tilde{t}_{1}}$,$m_{\tilde{t}_{2}}$ & 6035, 7065 & 4052, 9017 & \red{1873}, 11948 & 4534, 5196 & 6633, 8771 & 5864, 7613 \\ \hline 
$m_{\tilde{d}_{1}}$,$m_{\tilde{d}_{2}}$ & 7854, 8814 & 9098, 14101 & 10080, 18591 & 6031, 7041 & 8836, 11304 & 8185, 9493 \\
$m_{\tilde{b}_{1}}$,$m_{\tilde{b}_{2}}$ & 6037, 7621 & 4050, 9465 & 1869, 12879 & 4539, 5596 & 6634, 9494 & 5867, 8105 \\ \hline 
$m_{\tilde{\nu}_{e}}$,$m_{\tilde{\nu}_{\tau}}$ & 3971, 4981 & 5419, 9107 & 4916, 10153 & 1224, 3124 & 3115, 5210 & 3009, 4428 \\
$m_{\tilde{e}_{1}}$,$m_{\tilde{e}_{2}}$ & 4978, 6444 & 9100, 14155 & 10142, 18617 & 3122, 4931 & 5206, 8864 & 4424, 6894 \\
$m_{\tilde{\tau}_{1}}$,$m_{\tilde{\tau}_{2}}$ & 3971, 4706 & 5419, 8903 & 4916, 11641 & \red{1219}, 2497 & 3115, 6198 & 3011, 5357 \\ \hline 
$\sigma_{{\rm SI}}$ & $ 2.22 \times 10^{-12} $ & $ 4 \times 10^{-15} $ & $ 2 \times 10^{-15} $ & $ 2.04 \times 10^{-12} $ & $ 3.01 \times 10^{-12} $ & $ 2.82 \times 10^{-11} $ \\
$\sigma_{{\rm SD}}$ & $ 2.78 \times 10^{-8} $ & $ 1.2 \times 10^{-11} $ & $ 5.2 \times 10^{-12} $ & $ 3.91 \times 10^{-9} $ & $ 2.68 \times 10^{-9} $ & $ 7.27 \times 10^{-8} $ \\
$\Omega h^{2}$ & 0.121 & 0.121 & 0.12 & 0.122 & 0.115 & 0.117 \\ \hline 
$C$ & 0.107 & 0.117 & 0.155 & 0.17 & 0.177 & 0.093 \\ \hline 
\end{tabular}}
\caption{Benchmark points satisfying the mass bounds, the constraints from rare B-meson decays, Planck bounds within $5\sigma$ and the QYU condition with $\mid C \mid \leq0.2$. All masses are given in GeV. Point 1 represents chargino-neutralino coannihilation, and Point 2 exemplifies solutions in which the DM relic density is satisfied entirely through the gluino-neutralino coannihilation processes, while Point 3 depicts solutions for the stop-neutralino coannihilation scenario. Point 4 displays solutions for stau-neutralino coannihilation scenario. Points 5 and 6 represent the solutions for Wino-like and Higgsino-like LSP neutralino solutions respectively.}
\label{tab1}
\end{table}

Finally, before concluding, we present six benchmark points in Table \ref{tab1} which exemplify our findings. All points are selected to be consistent with the constraints applied in our analyses together with the QYU condition. All masses are given in GeV, and the DM scattering cross-sections are in pb. We show the NLSP species in red, and the masses of particles, which are relevant to the discussion in bold. Point 1 represents Bino-like LSP neutralino solutions whose spin-independent scattering cross-section lies just below the current exclusion bound provided by the XENON1T experiment, and it is expected to be tested very soon. These solutions will also be tested by Darwin. Since the spectrum involves heavy SUSY scalars, the relic abundance of LSP neutralino for such solutions can be reconciled with the Planck measurements through chargino-neutralino coannihilation processes. Point 2 exemplifies solutions in which the DM relic density is satisfied entirely through gluino-neutralino coannihilation processes with $m_{\tilde{g}}\gtrsim 1.3$ TeV. We exemplify the stop-neutralino coannihilation scenario with Point 3. Points 2 and 3 also represent Bino-like LSP solutions, but in comparison with Point 1, they lead to small spin-independent scattering cross-sections which fall below the neutrino floor and need more statistics to be tested in the direct detection DM experiments. Point 4 displays a stau-neutralino coannihilation scenario, and we observe that this is accompanied by chargino-neutralino coannihilation processes in order to achieve the desired dark matter relic abundance. Even though the LSP is Bino-like as in the previous points, it should be testable in direct detection experiments in the near future. Points 5 and 6 depict solutions for Wino-like and Higgsino-Like LSP solutions, with masses greater than about 1.5 TeV, and 1 TeV respectively. Point 6 also depicts a spectrum in which the CP-odd Higgs boson mass is about 2.5 TeV, which should be testable at HL-LHC through $A,H\rightarrow \tau\tau$ events.

\section{Conclusion}
\label{sec:conc}
We have explored the predictions for sparticle masses including dark matter and NLSP candidates in the framework of a supersymmetric $SU(4)_c \times SU(2)_L \times SU(2)_R$ model which incorporates third family quasi-Yukawa unification. An unbroken $Z_2$ gauge symmetry contained in the $422$ model acts as matter parity and ensures the presence of a viable neutralino dark matter candidate. Our solutions contain Bino-like, Wino-like and Higgsino-like DM solutions accompanied by a variety of NLSP candidates including gluino, stop, stau and chargino. The NLSP gluino can be as light as 1.3 TeV, which should be accessible at the LHC Run 3, while stop-neutralino coannihilations become relevant for $1.8 \lesssim m_{\tilde{t}_{1}}\simeq m_{\tilde{\chi}_{1}^{0}}\lesssim 2.3$ TeV. The NLSP slepton masses are in the $0.9-3$ TeV range, and these solutions are also involved in chargino-neutralino coannihilation processes in order to realize a consistent relic density of LSP neutralino. The Wino-like and Higgsino-like LSP neutralino solutions are associated with masses of order 1.5 and 1 TeV respectively. There also exist A-resonance solutions with the $m_A$ mass varying between 0.5 and 2.5 TeV. In this context, the A-resonance solutions can also be tested through the decay channel $A,H \longrightarrow \tau \tau$, which currently excludes solutions with $m_A$ $\lesssim$ 2 TeV in the large $tan\beta$ region. We display several benchmark points that highlight these solutions and show that the dark matter neutralino may be accessible in collider and other dark matter searches.

\section*{Acknowledgment}
AT is partially supported by the Bartol Research Institute, University of Delaware. The research of CSU is supported in part by the Spanish MICINN, under grant PID2019-107844GB-C22. CSU would like to thank Instituto F\'{i}sica The\'{o}rica of Universidad Aut\'{o}noma de Madrid, where part of his research has been conducted. We acknowledge Information Technologies (IT) resources at the University of Delaware, specifically the high performance computing resources for the calculation of results presented in this paper. CSU also acknowledges the resources supporting this work in part were provided by the CEAFMC and Universidad de Huelva High Performance Computer (HPC@UHU) located in the Campus Universitario el Carmen and funded by FEDER/MINECO project UNHU-15CE-2848.

\bibliographystyle{JHEP}
\bibliography{QYU.bib}

\end{document}